Superstructure, sodium ordering and antiferromagnetism in $Na_xCoO_2$ (0.75≤x≤1.0)

Y.G. Shi*, H. C. Yu, C. J. Nie, and J.Q. Li

Institute of Physics, Chinese Academy of Sciences, Beijing, People's Republic of China

Transmission-electron-microscopy investigation reveals the presence of a superstructure with a doubled period of $2d_{110}$ in the $Na_xCoO_2$ materials for x ranging from 0.75 to 1.0. Systematic analyses suggest that this superstructure in general appears just below a phase transition which could yield an anomalous kink in resistivity. We herein interpret this superstructure in terms of sodium ordering occurring at low temperatures. Measurements of magnetic susceptibility for all $Na_xCoO_2$ (0.75≤x≤1.0) materials show an antiferromagetic state with Neel temperature $T_N$~4K.

Author to whom correspondence should be addressed: syg@ssc.iphy.ac.cn

The $Na_xCo_2O_4$ system, because of its particular properties of large thermoelectric power coexisting with low electric resistivity, has attracted much attention in past several years [1–4]. Thereafter, the discovery of superconductivity of water-intercalated $Na_xCoO_2$ compound which is a breakthrough in the search for new layered transition metal oxide superconductors immediately spurred tremendous round of intense interest in this system [5,6]. Now in-depth study has uncovered a variety of novel electronic magnetic behavior which is thought to be engendered by strong interaction between the charge carriers in $Na_xCo_2O_4$ [7]. Recently, Foo et al. report a phase diagram of non-hydrated $Na_xCo_2O_4$ for 0.31<x< 0.75. It shows two distinct metallic states separated by an insulating state that is stabilized at x=0.5 by charge ordering of the holes and the Na ions [8]. Although great progress has been made, it will still be a long trip for people to interpret satisfactorily the essence of these novel electronic magnetic behaviors. More production of theoretical study and evidence of the experimental aspect are being expected. In this paper, we will report on the structural and physical properties as revealed from a series of $Na_xCoO$ samples.

The $Na_xCoO_2$ (0.45≤x≤1) samples in the present study were prepared by the conventional technique for oxide ceramics [9]. The raw materials were $Na_2CO_3$ (≥99.99%) and $Co_3O_4$(>99.9%) and weighed according to stern chemical ratios to produce the mixtures of the different component for $Na_xCoO_2$ (0.45≤x≤1). The mixed powders were calcined in the temperature range from 800 to 850°C in air for 8 to 12 hours. For the resistivity and magnetization measurements, the powders were pressed into pellets and sintered at 800°C in flowing $O_2$ for another 8 hours. Specimens for transmission-electron microscopy (TEM) observations were polished mechanically with a Gatan polisher to a thickness of around 50μm and then ion-milled by a Gatan-691 PIPS ion miller. In addition, we also prepared some thin samples for electron diffraction experiments simply by crushing the bulk material into fine fragments, which were then supported by a copper grid coated with a thin carbon film. The TEM investigations were performed on a H-9000NA TEM operating at the voltage of 300kV and a Tecnai F20 (200 kV) electron microscope with an atomic resolution of about 0.23nm. In order to observe the structural phase transition and to minimize the radiation damage under electron beam, we have performed our major TEM observations at low temperatures.

Structural measurements by means of x-ray diffraction and TEM observation have been performed on the samples of $Na_xCoO_2$ ($0.45 \leq x \leq 1.0$), the results indicate that samples with $0.6 \leq x \leq 1.0$ can be considered as single-phase materials that have an hexagonal unit cell. However, the samples with $x<0.57$ is not of single phase, but containing some other minor impurities, such as $Co_3O_4$ and $CoCO_3$ as observed. Figure 1(a) and (b) show the electron-diffraction patterns for the x=0.75 sample. They were taken at room temperature along the [010] and [001] zone-axis directions, respectively. All diffraction spots can be well indexed by a hexagonal unit cell with lattice parameters of a=0.28nm, and c=1.092nm, in agreement with the results of x-ray diffraction.

Structural modifications upon either the variation of doping levels (x) or the decrease of temperature have been systematically analyzed. A remarkable superstructure in $Na_{1-x}CoO_2$ ($0.75 \leq x \leq 1.0$) material has been observed at low temperatures. Occasionally, this superstructure could become visible at room temperature in some regions in the $NaCoO_2$ samples, but in general showing up as very weak defused spots in the diffraction patterns. Figure 1(c) and (d) show the electron diffraction patterns of the x=0.75 sample at the temperature of about 100K, illustrating the presence of clear superstructure reflections on the relevant planes of reciprocal space. In additional to the main reflections, the superstructure spots appear at the systematic (h+1/2, k, l) positions. It is noted that the superstructure spots are generally very sharp, directly indicating the long-coherent nature (>100nm) of the ordered state. Frequently, three or two sets of satellite reflections become visible around each basic Bragg spot as shown in Fig. 1(c), those are considered to originate from domains where the superstructure modulation wave vectors are rotated by $60^o$ with respect to one another. On the other hand, in accordance with the extinction rules for space group of $P6_3/mmc$, the main diffraction spots of (0, 0, 2n+1) should be absent in diffraction patterns, therefore, the presence of these spots in the experimental results, as shown Fig. 1b and Fig. 1d, is believed to arise from to double diffraction effects or local structural distortion in association with superstructure.

A better and clear view of the atomic structural feature of the $Na_xCoO_2$ crystal has been obtained by high-resolution TEM investigations in several typical materials. Fig. 2a shows a high-resolution electron macrograph of a $Na_{0.75}CoO_2$ crystal taken along the <001> direction. This high-resolution image was obtained from a thin region of a crystal; therefore, we expect that the Co-atom positions could be identified as the dark dots. On the other hand, because the Na and O could only give very weak contrast, it is hardly surprising that this experimental image has not yielded the identifiable contrasts in the shown image. Figure 2 (b) shows a high-resolution electron micrograph of a $NaCoO_2$ crystal taken at low temperature of around 100K , illustrating the superstructure fringe with the space of $2d_{100}$ as indicated by arrows. It is commonly observed that in most areas the superstructure occurs in one direction, this suggests that the superstructure in present system is essentially one-dimensional. Areas with clear superstructure frequently show up complex domain structure corresponding to different orientation variants.

Delmas et al. [10] reported that the $Na_xCoO_2$ system has several distinct phases, the nonstoichiometric $Na_{0.74}CoO_2$ material has the largest theoretical energy density when incorporated into electrochemical cells. Fig. 3a shows a schematic structural model for $Na_{0.75}CoO_2$. Ronald et al has characterized the structure of this phase by using neutron diffraction data, their structural refinement suggests sodium ions occupy two available sites unequally [10], the refined data show that the sodium ion occupies the (2/3,1/3,1/4) site about half the time (or is half full all the time) and the (0,0,1/4) site about 1/4 of the time, and the (1/3, 2/3, 1/4) site is not used at all. This means that the sodium ion prefers to occupy the position without the cobalt ions above, i.e. the (2/3, 1/3, 1/4) site, which has an experimental occupancy factor of 0.5. $Na^+$ ion on (0, 0, 1/4) site with $Co^{3+}$ ions above and below has a smaller occupancy factor of 2.3. Furthermore, on this site, bottleneck radius distance between the faces of oxygen ions is 0.81Å, which allows that the sodium ion move freely through this material as there is a tunnel available for motion between oxygen sheets, as discussed by Ronald et al in ref.10.

Electric conductivity analysis indicates the resistivity of $Na_xCoO_2$ materials are partially in correlation with Na ionic conduction. At room temperature, electronic conductivity is higher than ionic conductivity which is expected to decreases rapidly with lowering temperature. Our resistivity measurements suggest there is clear anomaly in the resistivity at the temperature of around 230K as discussed below. This anomaly is likely to be in connection with the formation of the superstructure. Hence, we attribute the superstructure in $Na_xCoO_2$ materials to Na ordering presumably occurring at low temperatures.

Figure 3(b) schematically illustrates a structural model for the ordered arrangements of the Na atoms occupied (2/3, 1/3, z ) positions, In order to simplify the drawing, the oxygen and the Co atoms are omitted. Our model considers that the sodium ions at (2/3,1/3, z ) have an occupy factor of 0.5 for x=0.75 sample, which could simply give an order structure with a doubled space superstructure. Sodium ions on the (0,0,1/4) site with an occupy factor of 0.23 have a large freedom and moving space [10], the additional Na ions in materials with x>0.75 should mainly occupy on the site. Actually, the changes of this superstructure along with the nominal Na-concentration (x) have been carefully examined at low temperatures. It is found that the superstructure is clearly visible in the range of 0.75<x<1.0 at low temperatures (<200K) with an unchangeable modulation wave vector, i.e. $q$=(1/2,0,0) ~1/2$d_{100}$.

We now go on to discuss the electric transport and magnetic properties of $Na_xCoO_2$ (0.45≤ x ≤1.0). All samples show metallic behavior (i.e. dρ/dT>0) in whole temperature range as reported in previous works[11]. Figure 4 shows the resistivity curves of $Na_xCoO_2$ materials with x=0.75 and x=1.0, respectively. The remarkable upturns in the resistivity curves can be clearly recognized at the temperature 240K for =0.75 and 270K for x= 1.0, as clearly shown in the insets. Our measurements show this anomalous temperature change slowly with Na concentration. We propose that this kind of anomalies arise from Na ordering which occurs at low temperatures. Previously, this kind of anomalies has also been noted by Motohashi et al.[11] who attributed this phenomenon to a structural transition.

The behaviors of the resistivity for $Na_xCoO_2$ (0.75≤ x ≤1.0) reveal certain interesting properties, For instance, as x=0.75, $\rho$ is linear below 40K and above 60K, and $\rho$ is proportional to $T^p$ in the two regions. We estimated p to be 0.42 below 40K and to be 0.86 above 60K. As x=1.0, $\rho$ is linear below 60K, from 100K to 250K and from 270K to 280K, and $\rho$ is proportional to $T^p$ in the three regions. We estimated p to be 0.38 below 60K, to be 0.68 from 100K to 250K and to be -0.67 above 270K. Systematic analyses on materials with x=0.55 revealed a variety of new properties, such as electron crystals and phonon glasses.

It is reported that $Na_xCoO_2$ materials with x<0.75 show paramagnetic behavior in the whole temperature range (4-300K) [8-11]. Recently, unconventional magnetic properties and transport behavior has been discussed in x=0.75 sample [11]. Bayrakci et al [12] reported a antiferromagnetic state with in Neel temperature $T_N$=19.8K in the $Na_{0.8}coO_2$ single crystal. The ordered moment is perpendicular to the $CoO_2$ sheets. Our measurements on materials with x ranging from 0.47 and 1.0 demonstrate that the samples with x<0.7 have the conventional paramagnetic properties, and the antiferromagetic feature exist commonly in $Na_xCoO_2$ (0.75<x<1.0) materials that have the superstructure unit cell. Fig.5a-c shows the magnetic susceptibilities measured for samples with x=0.75, 0.85, and 1.0, respectively. The noticeable antiferromagnetism signal normally appears below 5K, and becomes much more distinctive with the increase of Na concentration (x). Fig.5c displays the magnetic susceptibility for x=1.0 sample showing a noticeable antiferromagnetic state, the Neel temperature in this case is about 4K. Detailed analyses of the physical properties along with this anti-ferromagnetic phase transition are in progress.

In summary, we have performed an extensively investigation on the structure, electric resistive and magnetic properties on the $Na_xCoO_2$ materials with x ranging from 0.75 to 1.0.

These materials have a superstructure unit cell with the modulation wave vector of $q=a^*/2$, the appearance this superstructure can be fundamentally understood by Na-ion ordering occurring at low temperatures. Measurements of the electric transport properties reveal evident anomalies in the resistivity curve in association with the Na ordering. Antiferromagnetism appears for all samples with $0.75<x<1.0$, and $NaCoO_2$ ($x=1$) in particular shows up a sharp antiferromagnetic transition at the temperature of $T_N=4K$.

Acknowledgments

We would like to thank Professor Mr. W. W. Huang and Miss. S. L. Jia for their assistance in preparing samples and measuring some physical properties. The work reported here was supported by the ''Hundreds of Talents'' program organized by the Chinese Academy of Sciences, P. R. China, and by the ''Outstanding Youth Fund (J. Q. Li)'' with Grant No. 10225415.

Figure captions

Figure 1 Electron diffraction patterns of (a), (b) at room temperature and (c), (d) at 100K.

Figure 2 High resolution TEM images showing the (a) hexagonal structure and (b) the superstructure.

Figure 3(a) Schematic structural model of $Na_{0.75}CoO_2$. (b) A simplified structural model showing the Na-ordering.

Figure 4 Resistivity of $Na_xCoO_2$ for x=0.75 and 1.0 plotted as a function of temperature. Insets show the resistivity curves with evident anomalies at the temperatures of 240K and 270K, respectively.

Figure 5 Magnetic susceptibility curves of $Na_xCoO_2$ with x=0.75, 0.85 and 1.0, showing the presence of a low temperature antiferromagnetic state.

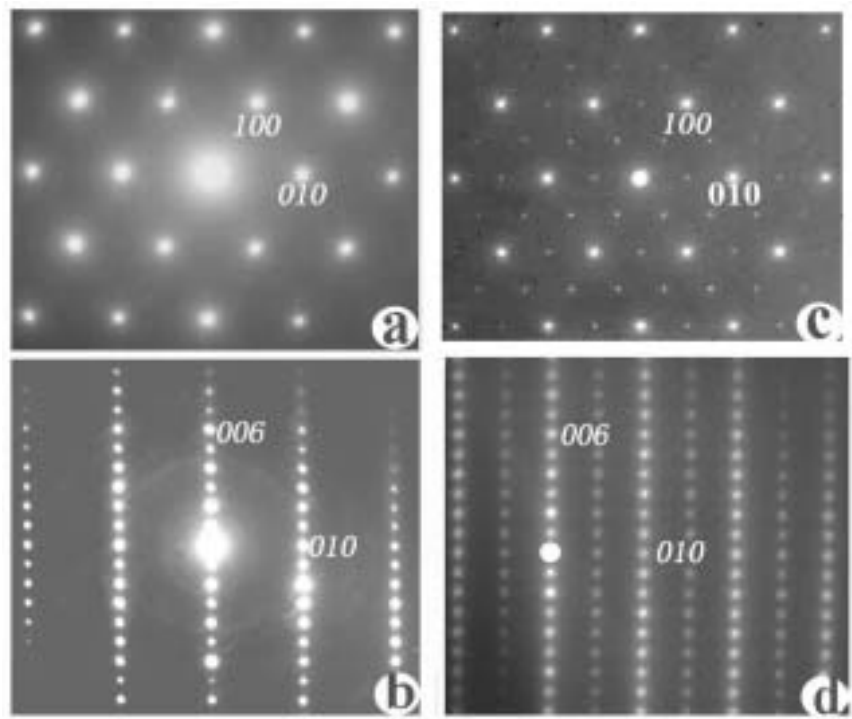

Figure 1

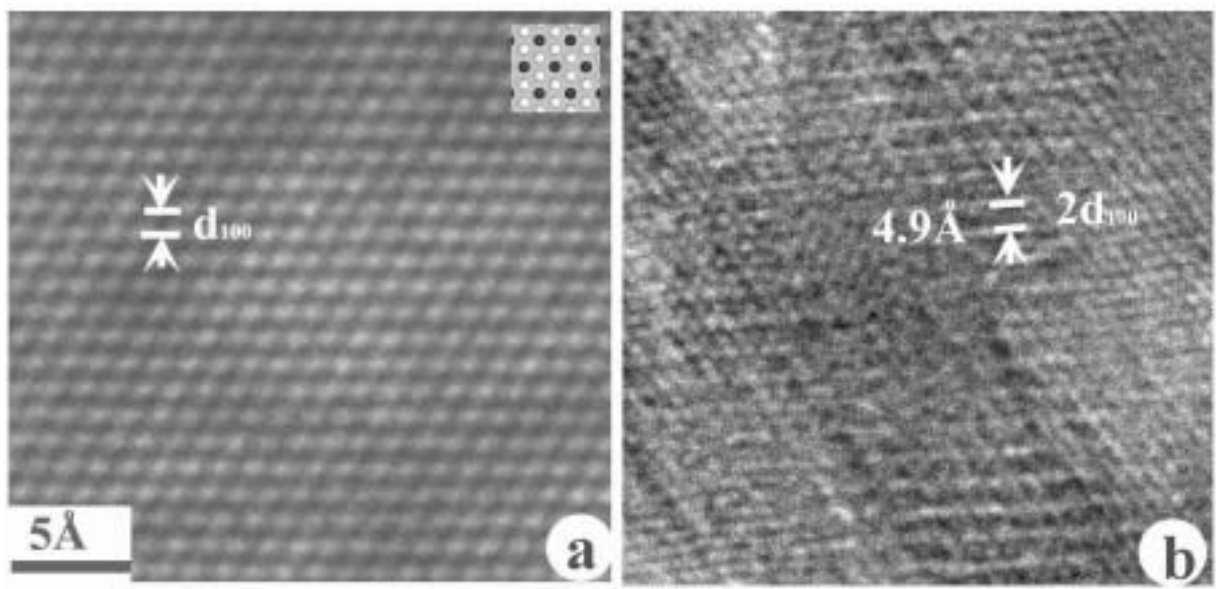

Figure 2

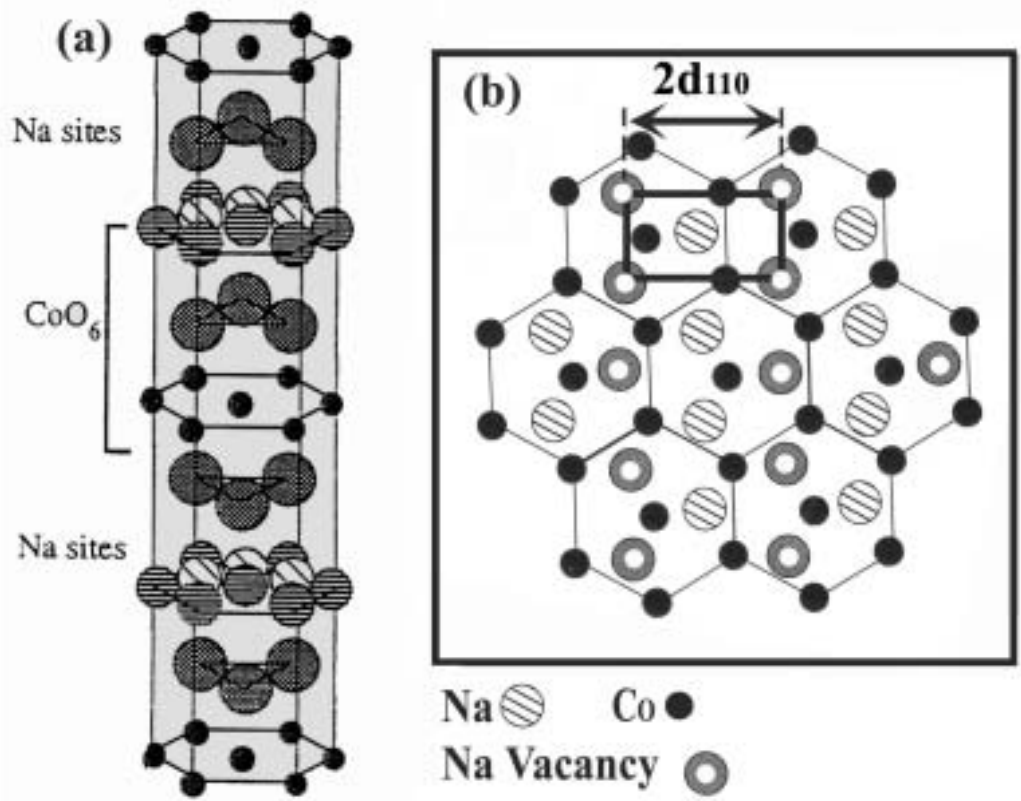

Figure 3

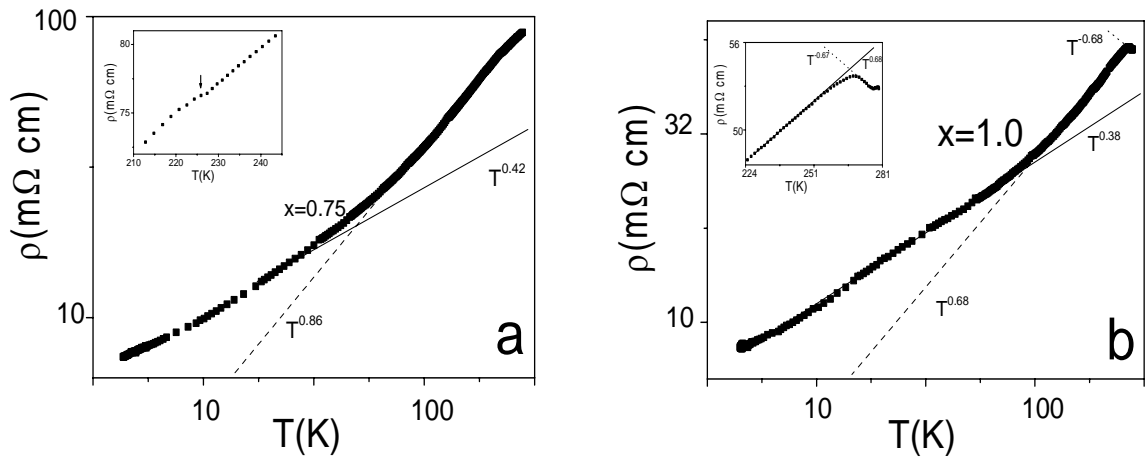

Figure 4

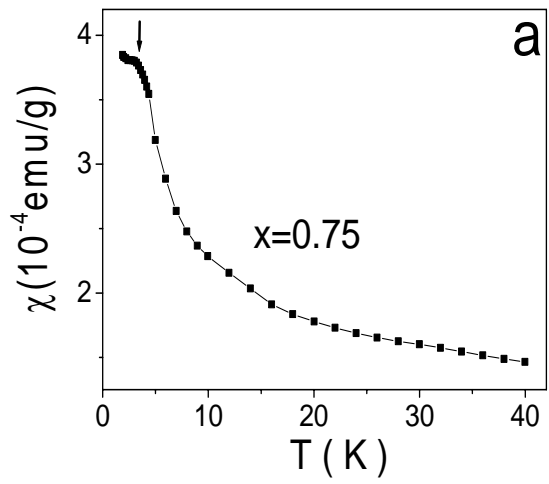
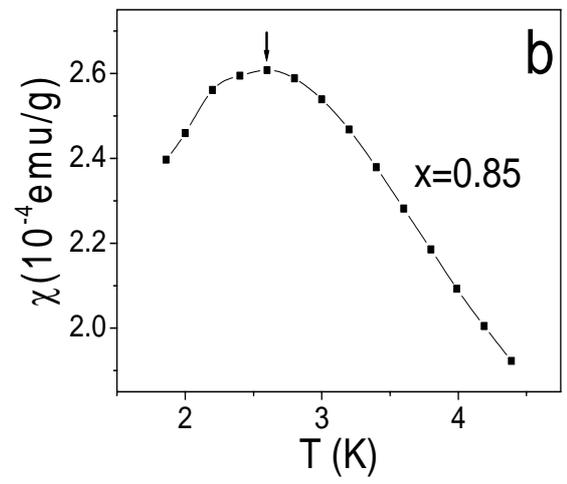
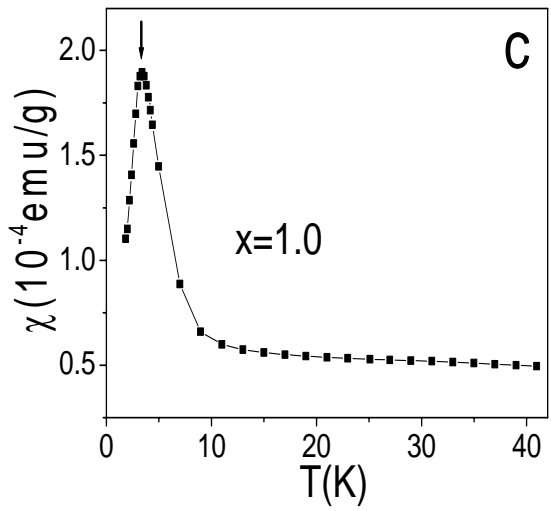

Figure 5